\documentclass[physics]{ejs_author}
\usepackage{upgreek}
\usepackage{color}
\usepackage{amsmath,amssymb,amsfonts,bbm,graphicx,graphics,times}
\usepackage[english]{babel}

\title{Quantum estimation of a two-phase spin rotation}

\articletype{Research Article}

\author{Cyril Vaneph\inst{1,2},
        Tommaso Tufarelli\inst{2},
        Marco G. Genoni\inst{2}\email{m.genoni@imperial.ac.uk}}

\institute{
     \inst{1} \'Ecole Normale Sup\'erieure de Lyon,\\ 15 parvis Descartes, BP 7000, \\
     69342 Lyon Cedex 07, France
     \inst{2} QOLS, Blackett Laboratory, Imperial College London,\\ London SW7 2BW, UK 
          }
\abr{QMTR}
\journal{Quantum Measurements and Quantum Metrology}

\abstract{We study the estimation of an infinitesimal rotation of a spin-$j$ system, characterized by two unknown phases, and compare the estimation precision achievable with two different strategies. The first is a standard `joint estimation' strategy, in which a single probe state is used to estimate both parameters, while the second is a `sequential' strategy in which the two phases are estimated separately, each on half of the total number of system copies.

In the limit of small angles we show that, although the joint estimation approach yields in general a better performance, the two strategies possess the same scaling of the total phase sensitivity with respect to the spin number $j$, namely $\Delta\Phi\simeq1/j$.

Finally, we discuss a simple estimation strategy based on spin squeezed states and spin measurements, and compare its performance with the ultimate limits to the estimation precision that we have derived above.}

\newcommand{\ket}[1]{|#1\rangle}
\newcommand{\bra}[1]{\langle#1|}
\newcommand{\ketbra}[2]{\vert #1 \rangle \! \langle #2 \vert}
\newcommand{\avr}[1]{\langle#1\rangle}

\def\Tr{\hbox{Tr}} 
\keywords{quantum metrology, multiparameter quantum estimation, spin squeezing}
\begin{document}
\firstpage{1}
\maketitle
\section{Introduction}
The rapidly developing field of {\it quantum metrology} studies how the peculiar features of quantum mechanics affect the achievable precision in the estimation of one or more parameters \cite{Giovanetti11}. The fundamental theoretical tools for these investigations are provided by {\it quantum estimation theory} \cite{Hel, BrauCaves, paris}, which gives fundamental bounds to the achievable accuracies in terms of the Quantum Fisher Information (QFI). Such theory is particularly useful when the parameters of interest are not accessible via a standard quantum-mechanical measurement. Indeed, most physicists are familiar with at least two problems of this type: the estimation of the optical phase (a {\it single parameter} estimation problem) and the experimental reconstruction of a system's density matrix (a {\it multiparameter} problem). In both cases, the parameters of interest cannot be measured directly, and they have to be inferred by a suitable post-processing of the measured data.

The quantum estimation of multiple parameters is receiving increasing attention in the literature (see for example \cite{Yuen,Hel,Belav, Holevo,Naga89,Fuji94,Ballester04,Chi06,You09,Wat10,MonIllu01,MonIllu02,Cro12,DispEst}). This kind of estimation is fundamentally different from the single-parameter case, as different parameters may be associated to mutually non-commuting operators. Hence, the theoretical bounds provided by local estimation theory may not be achievable, as they may assume measurement strategies that go beyond the operations allowed by quantum mechanics. In this context, remarkable efforts have been dedicated to the quantum estimation of a generic $SU(d)$ unitary operation \cite{Ballester04,Kahn07,Bisio10}. This is in line with the standard approach to Quantum Information Theory, which aims at characterizing the properties of quantum mechanical $d$-dimensional systems independently of their specific physical implementation. This general approach is certainly preferable if no assumptions are made regarding the structure of the system and of the unitary operations to be estimated. 

On the other hand, when some additional information {\it is} available regarding the physical nature of the setup at hand, it may be appropriate to place specific restrictions on the operations of interest. For example, suppose that the system in question is a physical spin-$j$ particle interacting with a classical magnetic field of fixed amplitude and direction. Then, the relevant operations to be estimated will belong to the rotation group $SO(3)$ (more precisely, to its spin-$j$ representation). In such a case, employing an estimation strategy aimed at the entire $SU(2j+1)$ group would be an unnecessary complication, if not a misuse of the available resources.

In this paper we are concerned with one such instance: we consider the quantum estimation of an infinitesimal rotation of a spin-$j$ system, characterized by two unknown phases. This may represent, for example, the result of the interaction between a spin-$j$ particle and a classical magnetic field lying in the $XY$ plane. We propose two different approaches: a `joint estimation' strategy, where a single probe state is used to estimate both parameters with a single measurement, and a `sequential estimation' strategy, where different optimized probe states 
are prepared, and the two phases are estimated 
separately, each on half of the number of system copies. We show that, though the two strategies give the same scaling in terms of the spin dimension $j$,  the joint estimation strategy yields in general a better performance. 

After having attacked the problem in its general form, we discuss the relevant special case in which the two-phase estimation is achieved by performing spin measurements on the probe states. Under this restriction, the concept of {\it spin squeezing} \cite{kita,vino} takes a central role in determining the performance of our estimation protocol. This agrees with the known results in the context of metrological tasks with spin systems. Indeed, instruments such as Ramsey spectrometers \cite{vino, precision, ramsey}, atomic clocks \cite{vino, entanglo2} and ultra-sensitive magnetometers \cite{mignot} have been shown to benefit from the presence of spin squeezing. We recall that, also in the context of quantum metrology with bosonic systems, squeezing is well established as being a desirable feature \cite{grave,busone}, able to enhance the performance of single-parameter \cite{phaseMon,Gaiba,lossMon, carmenKerr,OliBayes,PhestCV}, as well as multiparameter estimation protocols \cite{MonIllu01,MonIllu02,DispMatteo,DispEst}. 

The paper is organized as follows. In the next section we provide a brief introduction to local quantum estimation theory, both for the single- and the multi-parameter case. In Sec. \ref{s:SS} we
introduce the concept of spin squeezing, giving its definition for both the single- and the two-mode case. In Sec. \ref{s:TwoPh} we discuss in detail the estimation problem at the center of our investigations. After having described in detail the joint and sequential estimation strategies in their most general form, and the corresponding estimation precisions achievable, we discuss the role of spin squeezing for estimation protocols that rely on spin measurements. Sec. \ref{s:concl} concludes the paper with some remarks.
\section{Quantum estimation theory} \label{s:QET}
We will give here a brief review of quantum estimation theory \cite{Hel,paris,BrauCaves}.
Let us consider a family of quantum states $\varrho_\lambda$, labelled by a real
parameter $\lambda$ that we aim to estimate. 
The ultimate limit to the estimation precision of the parameter $\lambda$
is given by the quantum Cram\`er-Rao bound (QCRB)
\begin{align}
{\rm Var}(\lambda) \geq \frac{1}{M H(\lambda)},\label{singlebound}
\end{align}
where $M$ is the number of measurements performed, 
\begin{align}
H(\lambda) = {\rm Tr}[\varrho_\lambda L_\lambda^2] 
\end{align}
is the quantum Fisher information and $L_\lambda$ is the Symmetric Logarithmic Derivative (SLD) , that is, the operator satisfying
\begin{align}
2 \partial_\lambda \varrho_\lambda = L_\lambda \varrho_\lambda + \varrho_\lambda L_\lambda.
\end{align}
Note that, in principle, one can always find a quantum measurement able to attain equality in Eq.~\eqref{singlebound}, hence saturating the quantum Cram\`er-Rao bound.\\
Moving on to a multiparameter scenario, let us consider a family of quantum states $\varrho_{\bf z}$ labelled by
$d$ different parameters ${\bf z} = \{ z_\mu \}$, $\mu = 1, \dots, d$. The 
SLD for each parameter is defined via
\begin{align}
\frac{\partial\varrho_{\bf z}}{\partial z_\mu} &= \frac{L_\mu^{(S)} \varrho_{\bf z} + \varrho_{\bf z} L_\mu^{(S)}}{2}\, ,  \label{eq:SLD} 
\end{align}
from which one can calculate the QFI matrix
${\bf H}$:
\begin{align}
{\bf H}_{\mu\nu} &=  \Tr \left[ \varrho_{\bf z}\frac{ L^{(S)}_\mu L^{(S)}_\nu + L^{(S)}_\nu L^{(S)}_\mu}{2} \right ].
\end{align}
We define the covariance matrix elements $V({\bf z})_{\mu\nu}
= E[z_\mu z_\nu ] - E[ z_\mu] E[ z_\nu]$ and consider
a weight (positive definite) matrix ${\bf G}$. Then, the multiparameter
quantum Cram\'er-Rao bounds read
\begin{align}
{\rm tr} [ {\bf G V} ]   &\geq \frac1M {\rm tr}[ {\bf G} ({\bf H})^{-1}] \:,
\end{align}
where ${\rm tr}[A]$ is the trace operation on a finite dimensional matrix
$A$ and $M$ is the number of measurements performed. 
We observe that if we choose ${\bf G}=\mathbbm{1}$ we obtain
the bound on the sum of the variances of the parameters involved,
\begin{align}
\sum_\mu {\rm Var} (z_\mu) := \frac{(\Delta {\bf z})^2}M \geq \frac1M {\rm tr}[{\bf H}^{-1}]  
\label{eq:multiCRB}
\end{align}
where we have introduced the {\em overall} multiparameter sensitivity $\Delta {\bf z}$.
Differently from the single parameter case, the multiparameter bound is not always
achievable, since optimal measurements for different parameters may correspond
to non-commuting observables. 
A {\it sufficient} condition for the achievability of the Cram\'er-Rao bound is that the SLDs corresponding to different parameters commute on average on the probe state:
\begin{align}
	\Tr \left[ \varrho_{\bf z} [L^{(S)}_\mu, L^{(S)}_\nu ] \right ]=0.\label{com-av}
\end{align}
We remark that other (not always achievable) bounds to the estimation precision have been introduced in the multiparameter case, together with the concept of {\em most informative} 
bound \cite{Yuen,Belav,Holevo,Naga89,Fuji94}. In this manuscript we shall focus on the standard Cram\`er-rao bound described above.
\subsection{The pure state model}\label{sburo}
An important special class of quantum estimation problems is that in which the probe states are pure for all values of the parameters ${\bf z}$: $\varrho_{\bf z}=\ketbra{\psi_{\bf z}}{\psi_{\bf z}}$. Then, the SLDs are easily calculated as \cite{Ballester04}
\begin{align}
	L_\mu^{(S)}&=2\partial_\mu \varrho_{\bf z}.\label{SLDeasy}
\end{align}
It is convenient to introduce the auxiliary vectors
\begin{align}
	\ket{l_\mu}&=L_\mu^{(S)}\ket{\psi_{\bf z}},\label{aux}
\end{align}
in terms of which the QFI matrix elements simplify to
\begin{align}
	{\bf H}_{\mu\nu}&={\rm Re}\left[\langle l_{\mu}|l_\nu\rangle\right].\label{QFIeasy}
\end{align}
Remarkably the sufficient condition for the achievability of the QCRB, as given in Eq.~\eqref{com-av}, becomes here also necessary \cite{cazzumoto}. This can be re-expressed as follows: an estimation strategy saturating the bound exists if and only if
\begin{align}
	{\rm Im}\left[\langle l_{\mu}|l_\nu\rangle\right]=0.\label{cond-easy}
\end{align}
The pure state model is particularly relevant when the parameters ${\bf z}$ to be estimated are associated to a family of unitary maps. Indeed, suppose that the probe states can be expressed as $\varrho_{\bf z}=U({\bf z})\varrho_0U({\bf z})^\dagger$, where $\varrho_0$ is a generic mixed state of a Hilbert space $\mathcal H_0$, and $U({\bf z})$ is a unitary family on the same space. Let us now consider a generic purification of $\varrho_0$, i.e., a pure state $\ket{\psi_0}\in\mathcal H_0\otimes\mathcal H_A$, where $\mathcal H_A$ is an ancillary Hilbert space, such that $\Tr_A\left[\ketbra{\psi_0}{\psi_0}\right]=\varrho_0$. Since any measurement strategy possible in the space $\mathcal H_0$ is also possible in $\mathcal H_0\otimes\mathcal H_A$, but not vice-versa, one has that the family of pure states $\ket{\psi_{\bf z}}=U({\bf z})\otimes\mathbb I_A\ket{\psi_0}$ in general allows a lower (or equal) QCRB for the parameters ${\bf z}$, with respect to the mixed states $\varrho_{\bf z}$.
\section{Spin systems and spin squeezing}\label{s:SS}
Let us consider a spin-$j$ system, characterized by the angular momentum operators
$\hat{\vec J}=( \hat{J}_x, \hat{J}_y,\hat{J}_z)^{\sf T}$, which satisfy the commutation rule
\begin{align}
[\hat{J}_\alpha, \hat{J}_\beta] = i \hat{J}_\gamma.
\end{align}
where $\alpha$, $\beta$, $\gamma$ are a right-handed permutation of $x,y,z$.
This implies an uncertainty relation
\begin{align}
\Delta \hat{J}_x \Delta\hat{J}_y \geq \frac {|\langle \hat{J}_z \rangle |}{2}, \label{eq:uncrel}
\end{align}
where $\Delta \hat{J}_\alpha^2 = \langle \hat{J}_\alpha^2\rangle - \langle \hat{J}_\alpha\rangle^2$
and $\langle O \rangle = \Tr[\varrho O]$ denotes the expectation value of 
an operator on a given quantum state $\varrho$.  Several definitions of spin squeezing have been introduced, the most commonly adopted being that suggested by Kitagawa and Ueda \cite{kita}, and by Wineland {\em et al.} \cite{vino}. In the following we will present the details regarding the latter definition, which is more relevant to metrological applications. For a generic spin state, the mean-spin direction (MSD) is defined as
\begin{align}
\vec{n} =\frac{\langle \hat{\vec{J}}\, \rangle }{|\langle \hat{\vec{J}}\, \rangle |}.
\end{align}
Without loss of generality, let us consider a state with MSD along the $z$-axis (in particular, $\avr{\hat J_x}=\avr{\hat J_y}=0$). Then, suppose the system undergoes an infinitesimal rotation of angle $\phi$ around the $x$-axis. In the Heisenberg picture, one has, 
\begin{align}
\hat J_y\to\hat{J}_y^{\sf out}\simeq \hat J_y+\phi\hat J_z.\label{easyrot}
\end{align}
The above relations may then be exploited to estimate the value of $\phi$ from the measurement statistics of the observable $\hat{J}_y^{\sf out}$, via $\avr{\hat{J}_y^{\sf out}}=\phi\avr{\hat J_z}$. One can see that the `phase sensitivity' corresponding to this estimation strategy is
\begin{align}
\Delta \phi = \frac{\Delta \hat{J}_y^{\sf out}}{|\partial_\phi \langle \hat{J}_y^{\sf out} \rangle |} 
\approx \frac{ \Delta \hat{J}_y }{|\langle \hat{J}_z \rangle |},\label{sensi-spin}
\end{align}
and is thus related to the initial variance of the operator $\hat{J}_y$. It is useful for our purposes to introduce the notion of coherent spin states (CSS). Given a unit vector $\vec n$, these are defined as 
\begin{align}
|C(\vec n)\rangle = |j,j\rangle_{\vec n}
\end{align}
where $|j,j\rangle_{\vec n}$ is the eigenstate of $\hat{J}_{\vec n}=\vec n\cdot \hat{\vec J}$ with eigenvalue $j$. For a coherent spin state, the fluctuations of all the spin operators orthogonal to the the MSD are equal to $(\Delta \hat{J}_{\vec{n}_\bot})^2 = j/2$ and then one proves that
\begin{align}
\Delta \phi_{\sf SQL}= \frac{1}{\sqrt{2 j}}, \label{eq:SQL}
\end{align}
which is referred to as the standard quantum limit (SQL) or shot-noise limit. Just as in the bosonic case, the coherent states provide a reference for the definition of squeezing. Adopting the definition of Ref.~\cite{vino}, a state is said to be spin squeezed iff its corresponding phase sensitivity is below the shot noise limit
\begin{align}
\Delta \phi \leq \Delta\phi_{\sf SQL},
\end{align}
that is, if and only if the state provides a better phase sensitivity as compared to a CSS. 

One can extend the concept of squeezing to a multipartite system. Let us briefly review two-mode spin squeezing \cite{Kuz00,Jul01,Ber02PRA,Ber02NJP,Ray03}, for a bipartite spin system
described by operators $\hat{\vec{J}}_a =( \hat{J}_{x_a}, \hat{J}_{y_a},\hat{J}_{z_a})^{\sf T}$
with $a=\{1,2\}$. Sums and differences of spin operators
belonging to different systems are denoted by $\hat{J}_{\alpha \pm} = \hat{J}_{\alpha_1}
\pm J_{\alpha_2}$, and obey the uncertainty relations
\begin{align}
	\Delta \hat{J}_{x \pm}\Delta\hat{J}_{y \pm}\geq\frac{|\langle\hat{J}_{z+}\rangle|}{2}.\label{unc2}
\end{align}
A state is said to be two-mode spin squeezed if
\begin{align}
(\Delta \hat{J}_{x-})^2 + (\Delta \hat{J}_{y+})^2 \leq |\langle \hat{J}_{z+} \rangle | . \label{eq:twoSS}
\end{align}
Here, the reduced fluctuations are not observed in the local degrees of freedom, but rather in the non-local operators $\hat{J}_{x-}$ and $\hat{J}_{y+}$. In fact, two-mode spin squeezing is not only a sufficient \cite{Ber02NJP,Ray03}, but also a necessary condition for entanglement in pure states of two subsystems with equal spin (except
for a set of bipartite states with measure zero) \cite{Ber05}.
\section{Estimation of a two-phases spin rotation} \label{s:TwoPh}
In this paper we consider the following estimation problem: a probe state $\varrho$ undergoes
a unitary evolution
\begin{align}
\hat{U}(\vec \phi) = \exp \left\{ i \vec\phi\cdot\hat{\vec J}\right\}\label{U}, 
\end{align}
where $\vec \phi=(\phi_x,\phi_y,0)$, $\phi_x$ and $\phi_y$ are two unknown phases which we aim to estimate, while $\hat{\vec J}$ is a vector of spin-$j$ operators as above. In the case of a physical spin, the above rotation could be the result of the interaction between the system and a classical magnetic field lying in the XY plane. Since we shall be interested in studying the ultimate limits on the achievable precision of such an estimation, and the transformation \eqref{U} is unitary, we can restrict our attention to the pure state model as explained in Section~\ref{sburo}. 
Let us then consider a generic input state $\ket{\psi_0}\in\mathcal H_j\otimes\mathcal H_A$, where $\mathcal H_j$ is the Hilbert space of a spin-$j$ particle on which the transformation $\hat U$ is acting, while $\mathcal H_A$ is associated to a generic ancillary system. The evolved states are then
\begin{align}
	\ket{\psi_{\vec\phi}}=\hat U(\vec\phi)\ket{\psi_0}.\label{probes}
\end{align}
Notice that different figures of merit can be introduced to quantify the performance of a multi-parameter estimation strategy. When no assumption is made about the unitary operation to be estimated, a parametrization-independent figure of merit may be preferable (see e.g. \cite{Ballester04}). In our case, however, there is a particular parametrization of the unknown unitary, expressed by Eq. (\ref{U}), which has a clear physical interpretation in terms of spin directions and rotation angles. We thus consider a figure of merit for our estimation protocol based on this particular parametrization: namely the sum of the variances of the two estimated phases. This corresponds to the QCRB in Eq. (\ref{eq:multiCRB}) .\\
We are interested in the limit of infinitesimal rotations, i.e. $|\vec\phi|\!\ll\!1$, where Eq.~\eqref{U} can be linearized as
\begin{align}
	\hat U(\vec\phi)\simeq\mathbb I+i\vec\phi\cdot\hat{\vec J}.\label{Uapp}
\end{align}
Combining this with Eqs.~\eqref{SLDeasy} and \eqref{aux}, one can easily calculate the auxiliary vectors $\ket{l_\mu}$, up to first order in $\vec\phi$, according to
\begin{align}
	\ket{l_\mu}&\simeq\ket{l_\mu^{(0)}}+\ket{l_\mu^{(1)}},\\
	\ket{l_\mu^{(0)}}&=2i\left(\hat J_\mu\ket{\psi_0}-\ket{\psi_0}\bra{\psi_0}J_\mu\ket{\psi_0}\right),\label{l0}\\
	\ket{l_\mu^{(1)}}&=2\left(\vec\phi\cdot\hat{\vec J}\ket{\psi_0}\bra{\psi_0}\hat J_\mu\ket{\psi_0}-\ket{\psi_0}\bra{\psi_0}\hat J_\mu\vec\phi\cdot\hat{\vec J}\ket{\psi_0}\right).
\end{align}
Correspondingly, from Eq.~\eqref{QFIeasy} one is able to expand the QFI matrix elements as
\begin{align}
	{\bf H}_{\mu\nu}&\simeq{\bf H}_{\mu\nu}^{(0)}+{\bf H}_{\mu\nu}^{(1)},\label{Hexpanded}\\
	{\bf H}_{\mu\nu}^{(0)}&=4\avr{\Delta\hat J_\mu\hat J_\nu}_0=4\left[\tfrac{1}{2}\avr{\{\hat J_\mu,\hat J_\nu\}}_0-\avr{\hat J_\mu}_0\avr{\hat J_\nu}_0\right],\label{H0}\\
	{\bf H}_{\mu\nu}^{(1)}&=2i\left[\avr{[\vec\phi\cdot\hat{\vec J},\hat J_\nu]}_0\avr{\hat J_\mu}_0+\avr{[\vec\phi\cdot\hat{\vec J},\hat J_\mu]}_0\avr{\hat J_\nu}_0\right],\label{H1}
\end{align}
where $\avr{\hat A}_0=\bra{\psi_0}\hat A\ket{\psi_0}$. 
In what follows, we shall focus on the regime $|\vec\phi|\ll1$, in which the second term in Eq.~\eqref{Hexpanded} can be neglected, and the QFI matrix reduces to ${\bf H}_{\mu\nu}\simeq{\bf H}_{\mu\nu}^{(0)}$, that is, the covariance matrix of the operators $\hat J_x$ and $\hat J_y$, [see Eq.~\eqref{H0}]. Once the optimal quantum states for our estimation problem have been calculated in this regime, we may use the first-order correction in Eq.~\eqref{H1} to estimate the range of validity of  the zeroth-order approximation. We are now ready to illustrate the two estimation strategies at the center of our investigations.
\subsection{Joint estimation strategy}
\begin{figure}[h!]
\includegraphics[width=0.95\columnwidth]{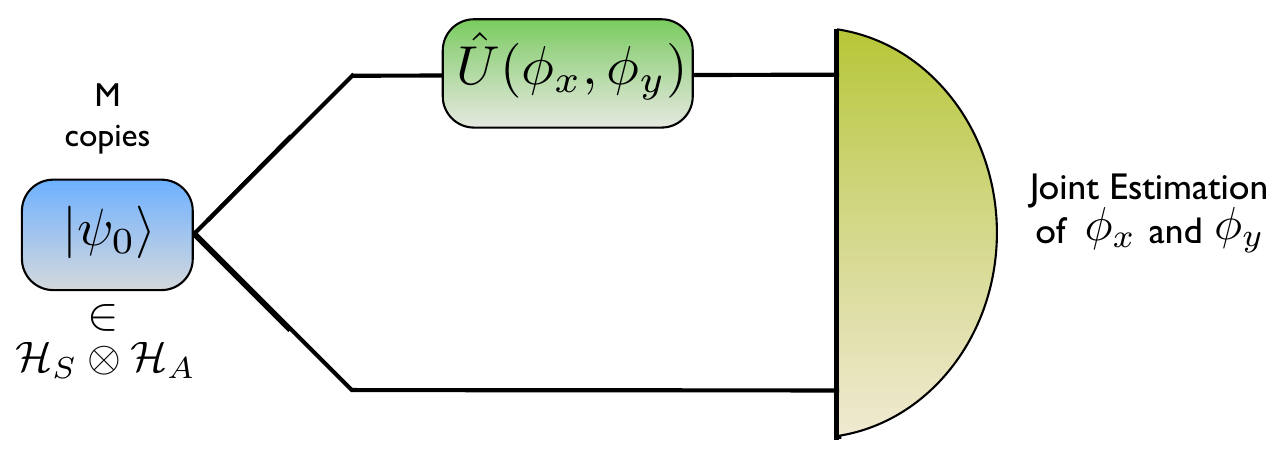}
\caption{`Joint estimation' strategy for a two-phases rotation. $M$ copies of an initial probe state $|\psi_0\rangle$ are prepared (in general this may involve an ancilla with which the system is entangled). After the action of the unitary operation $U(\phi_x,\phi_y)$, a joint measurement is performed on the system plus the ancilla, followed by post-processing for the determination of $\phi_x,\phi_y$.
\label{f:schemeJES}}
\end{figure}

We shall start by considering the standard approach to multiparameter estimation, and look for a single input state $\ket{\psi_0}$ which allows high sensitivity in the estimation both parameters $\phi_x,\phi_y$ as described in Fig. \ref{f:schemeJES}. 
According to the QCRB in Eq.~\eqref{eq:multiCRB}, this amounts to finding the quantum state minimizing the quantity
\begin{align}
\Tr[{\bf H}^{-1}]\simeq\Tr[({\bf H}^{(0)}
)^{-1}].
\end{align}
Exploiting Eq.~\eqref{H0}, and the fact that for any operator $\hat A$ one has $\avr{\Delta\hat A^2}_0\leq\avr{\hat A^2}_0$, we have
\begin{align}
\Tr[({\bf H}^{(0)}
)^{-1}]&=\frac{\avr{\Delta\hat J_x^2}_0+\avr{\Delta \hat J_y^2}_0}{4\left[\avr{\Delta\hat J_x^2}_0\avr{\Delta\hat J_y^2}_0-(\avr{\Delta\hat J_x\hat J_y}_0)^2\right]}\nonumber\\
&\geq\frac14\left(\frac{1}{\avr{\Delta\hat J_x^2}_0}+\frac{1}{\avr{\Delta\hat J_y^2}_0}\right)\nonumber\\
&\geq\frac14\left(\frac{1}{\avr{\hat J_x^2}_0}+\frac{1}{\avr{\hat J_y^2}_0}\right).\label{multi-ineq}
\end{align}
By minimising the expression \eqref{multi-ineq} under the constraint $\avr{\hat J_x^2}_0+\avr{\hat J_y^2}_0=\avr{\hat J^2}_0-\avr{\hat J_z^2}_0$ one obtains the lower bound
\begin{align}
\Tr[({\bf H}^{(0)})^{-1}]\geq\frac{1}{j(j+1)-\avr{\hat{J}_z^2}_0}.\label{jointbound}
\end{align}
The above derivation shows that saturating the inequality \eqref{jointbound} requires the necessary conditions
\begin{align}
\avr{\Delta\hat J_x\hat J_y}_0&=0\label{covazero},\\
\avr{\hat J_x}_0=\avr{\hat J_y}_0&=0,\label{zeromedia}\\
\avr{\hat J_x^2}_0=\avr{\hat J_y^2}_0&=\frac12\left[j(j+1)-\avr{\hat{J}_z^2}_0\right].\label{iso}
\end{align}
In order to discuss concrete limits to the estimation precision, we shall focus on \textit{achievable} bounds. We recall that a necessary and sufficient condition for the achievability of the QCRB is given by Eq.~\eqref{cond-easy}. With the use of Eq.~\eqref{l0} one obtains ${\rm Im}\left[\langle l_{\mu}|l_\nu\rangle\right]\propto\avr{[\hat J_\mu,\hat J_\nu]}$, hence in the present context the QCRB is achievable if and only if
 \begin{align}
 \avr{\hat J_z}_0=0.\label{zetazero}
 \end{align}
Our task is then to find an initial state verifying Eqs.~\eqref{covazero}, \eqref{zeromedia}, \eqref{iso} and \eqref{zetazero}, while at the same time achieving the lowest possible value of $\avr{\hat J_z^2}_0$. If $j$ is an integer, it is easy to check that the state
\begin{align}
\ket{\psi_0}=\ket{j,0}
\end{align}
is the optimal choice, since it satisfies all the above constraints, and yields $\avr{\hat J_z^2}_0=0$. Note that in this case we have not needed to introduce an ancillary system.
The case in which $j$ is semi-odd is slightly more complicated, since $\avr{\hat J_z^2}_0=0$ is not achievable. Guided by this consideration, we look for states which yield the minimum value of $\avr{\hat J_z^2}$, namely $1/4$. This, together with Eq.~\eqref{zetazero}, implies that our state must be an equally-weighted superposition of the states $\ket{j,\pm\tfrac12}$. Furthermore, in order to satisfy Eqs.~\eqref{covazero} and \eqref{zeromedia}, we shall make use of an ancilla with which our spin-$j$ system is entangled. We thus consider a state of the form
\begin{align}
\ket{\psi_0}=\frac{1}{\sqrt2}\left(\ket{j,\tfrac12}\ket{0}_A+{\rm e}^{i\varphi}\ket{j,\!-\!\tfrac12}\ket{1}_A\right),\label{seqstatex}
\end{align} 
Where $\ket0,\ket1$ are two orthonormal states in the ancillary Hilbert space. Easy calculations show that such a state verifies $\avr{\hat J_z}_0=0$ and $\Tr[({\bf H}^{(0)})^{-1}]=[j(j+1)-\frac14]^{-1}$ as required. To summarize, we have shown that the optimal phase sensitivity for the join estimation protocol, defined as $\Delta\Phi_{\sf JE}=\sqrt{\Tr[{\bf H^{-1}}]}$, is given by
\begin{align}
\Delta\Phi_{\sf JE}=\left\{\begin{array}{lcr}
\frac{1}{\sqrt{j(j+1)}}& \phantom{0}&j\text{ integer},\\
\frac{1}{\sqrt{j(j+1)-\frac14}}& \phantom{0}&j\text{ semi-odd}.
\end{array}\right. \label{eq:JES}
\end{align}
Remarkably, the optimization of the initial state yields very different results depending on whether $j$ is integer or semi-odd: one obtains an eigenstate of $\hat J_z$ in the former case, and an entangled system-ancilla state in the latter.
Finally, let us comment on the zeroth-order approximation of Eq.~\eqref{Hexpanded} that we have adopted. Since the optimal states that we have derived verify $\avr{\hat{\vec{J}}}_0=0$, one can see that the first order correction to the QFI matrix [Eq.~\eqref{H1}] is identically zero. Hence, our results are accurate up to second order in $|\vec\phi|$.

\subsection{Sequential estimation strategy}\label{s:seq}
\begin{figure}[h!]
\includegraphics[width=0.95\columnwidth]{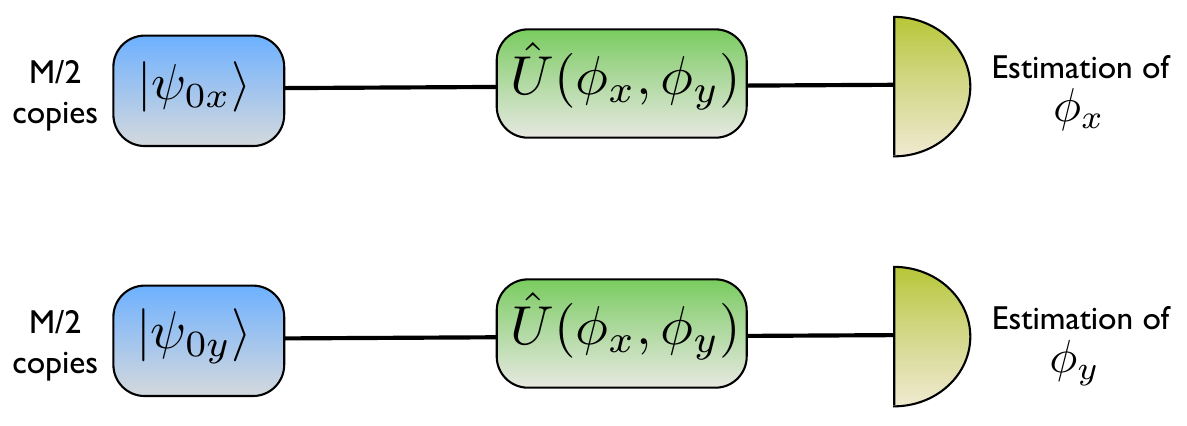}
\caption{`Sequential estimation' strategy for a two-phases measurement.
The total number $M$ of system copies is divided in two ensembles, such that $M/2$ copies are devoted to the estimation of $\phi_x$, and $M/2$ copies to that of $\phi_y$. The optimal initial states and measurement strategies for the two ensembles are in general different.
\label{f:schemeSES}}
\end{figure}

The next estimation scheme, sketched in Fig. \ref{f:schemeSES},  is based on the simple idea of combining two single-parameter estimation protocols. 
Given $M$ copies of the system, we partition them in two ensembles composed of $M/2$ copies each. Then, without loss of generality, we can assume that the first $M_x= M/2$ copies are used to estimate the phase $\phi_x$, while the remaining $M_y=M/2$ copies are dedicated to the estimation of $\phi_y$. Note that the optimization of such a procedure can be expected to provide different initial states and measurement strategies for the two ensembles. Let us now focus on the estimation of $\phi_x$ on the first ensemble. As we are not interested in the parameter $\phi_y$, the QFI for this case is simply given by the in-diagonal element of the QFI matrix
\begin{align}
H(\phi_x)={\bf H}_{xx}.
\end{align}
At zeroth order in $|\vec\phi|$, this is simply proportional to the variance of $\hat J_x$ on the initial state:
\begin{align}
{\bf H}_{xx}^{(0)} &=4\bra{\psi_0}\Delta \hat J_x^2\ket{\psi_0}\leq4j^2,\label{singlelimit}
\end{align}
yielding an asymptotic variance of the estimated phase
\begin{align}
{\rm Var}(\phi_x) &= \frac{\Delta\phi_x}{M_x} \geq \frac{1}{4M_x j^2}.\label{Hlimit}
\end{align}
Note that the inequality in Eq.~\eqref{singlelimit} can be saturated only if $\avr{\hat J_x}=0$ and $\avr{\hat J_x^2}=j^2$. One can verify that the only states verifying both conditions are of the form
\begin{align}
\ket{\psi_{0x}}=\tfrac{1}{\sqrt2}\left(\ket{j,j}_x+ e^{i\xi} \ket{j,-j}_x\right), \label{eq:optx}
\end{align}
where $\ket{j,m}_x$ are eigenstates of $\hat{J}_x$ with eigenvalue $m$. Since we are dealing with a single parameter estimation problem, the bound is always achievable by means of a projective measurement, and a suitable post-processing of the data. As regards the estimation of $\phi_y$, one still obtains at zeroth order
\begin{align}
H(\phi_y) &= {\bf H}_{yy} \leq 4j^2  \\
{\rm Var}(\phi_y) &= \frac{\Delta\phi_y}{M_y} \geq \frac{1}{4M_y j^2}.
\end{align}
which is saturated by a state 
\begin{align}
\ket{\psi_{0y}}=\tfrac{1}{\sqrt2}\left(\ket{j,j}_y+ e^{i\xi} \ket{j,-j}_y\right). \label{eq:opty}
\end{align}
Adopting a sequential strategy, and thus dividing the protocols in $M_x=M_y=M/2$ copies for the estimation of each of the two parameters, when $M\gg1$, the Central Limit Theorem predicts the total experimental variance of the estimated parameters, as per
\begin{align}
{\rm Var}(\phi_x)+{\rm Var}(\phi_y) \approx \frac{2}{M}\left(\Delta\phi_x^2+\Delta\phi_y^2\right).\label{var}
\end{align}
where $\Delta\phi_x^2\geq 1/(4j^2)$ and $\Delta\phi_y^2\geq 1/(4j^2)$ as derived above in the 
two single-parameters Cram\'er-Rao bounds.
A comparison of this expression with the multi-parameter Cram\`er-Rao bound in Eq. (\ref{eq:multiCRB}), which is also expressed in terms of the {\it total} number of measurements $M$, leads us naturally to define the {\em effective} phase sensitivities
\begin{align}
\Delta \phi_x^{\sf eff} = \sqrt{2} \Delta\phi_x \qquad \Delta \phi_y^{\sf eff} = \sqrt{2} \Delta\phi_y \:.
\label{eq:effps}
\end{align}
Then, the performance of our estimation protocol can be quantified via the two-phase sensitivity
\begin{align}
\Delta\Phi &\approx M [{\rm Var}(\phi_x)+{\rm Var}(\phi_y)] \nonumber \\
&=\sqrt{ (\Delta \phi_x^{\sf eff} )^2 + (\Delta \phi_y^{\sf eff} )^2}. \label{eq:phasesens}
\end{align}
By adopting the optimal states in Eqs. (\ref{eq:optx}) and (\ref{eq:opty}), one achieves the optimal phase-sensitivity for the sequential estimation strategy as per
\begin{align}
\Delta\Phi_{\sf SE} = \frac 1j . \label{eq:SES}
\end{align}
By comparing this result with the one in Eq. (\ref{eq:JES}), one observes that the same asymptotic scaling is obtained for large values of the spin dimension $j$, while, for intermediate values, the joint estimation strategy yields a better performance. Also in this case, we can see that the first order corrections ${\bf H}_{xx}^{(1)}$ and ${\bf H}_{yy}^{(1)}$ [Eq.~\eqref{H1}] are identically zero for the considered probe states [Eqs.~\eqref{eq:optx} and \eqref{eq:opty}], indicating that our treatment is valid up to second order in $|\vec\phi|$.
\section{Restricting to spin measurements}\label{s:spinmeas}
So far we have discussed general limits to the estimation precision, which assumed no restriction on the available measurement strategies. Depending on the specific physical implementation, it can be the case that only a restricted set of measurements can be performed on the probe states. This motivates us to study the simple but relevant special case in which only spin measurement are available. For this purpose, we find it convenient to treat the problem in the Heisenberg picture, that is, we shall keep fixed the probe state $\ket{\psi_0}$, and apply the rotation to the spin operators $\hat{\vec J}$, according to
\begin{align}
\hat{\vec J}^{\sf out}=\hat U(\vec \phi)^\dagger\,\hat{\vec J}\,\hat U(\vec \phi)\simeq \hat{\vec J}-i[\vec\phi\cdot\hat{\vec J},\hat{\vec J}].\label{Heisenberg}
\end{align}
In particular, for the $x$ and $y$ components one has
\begin{align}
\hat{J}^{\sf out}_{x}&\simeq\hat J_x-\phi_y\hat J_z,\label{spinphiy}\\
\hat{J}^{\sf out}_{y}&\simeq\hat J_y+\phi_x\hat J_z.\label{spinphix}
\end{align}
\subsection{Sequential strategy}
We start by discussing a `sequential estimation strategy' based on spin measurements. Following the same lines as in Section~\ref{s:seq}, we partition $M$ copies of the system in two ensembles of $M_x=M_y=M/2$ copies. The first $M_x$ copies shall be employed to estimate the phase $\phi_x$, by exploiting Eq.~\eqref{spinphix}. Following the same procedure outlined in Section~\ref{s:SS}, we find that the resulting phase sensitivity for the estimation of $\phi_x$ is
\begin{align}
\Delta\phi_x&\simeq\frac{ \Delta \hat{J}_y }{|\langle \hat{J}_z \rangle |}.\label{spinsensix}
\end{align}
Hence, we find also in this case that the coherent spin states provide the standard quantum limit $\Delta\phi_x^{\sf SQL}=1/\sqrt{2j}$. To improve this figure, spin-squeezed states are required. We note at this point that the optimal scaling $\Delta\phi_x=1/2j$, also known as the {\it Heisenberg limit}, is not achievable within the present context. We have indeed shown in Section~\ref{s:seq} that the only states able to achieve such precision are those given in Eq.~\eqref{seqstatex}. However, these states yield $\avr{\hat J_z}=0$, meaning that they cannot be used to estimate the value of $\phi_x$ via Eq.~\eqref{spinphix} [indeed, the corresponding variance in Eq.~\eqref{spinsensix} would diverge]. Nevertheless, it is still possible to find spin-squeezed states providing the same asymptotical scaling $\Delta\phi_x\propto1/j$, as shown numerically in Ref.~\cite{kita}, and analytically in Ref.~\cite{yurke} . We provide here a constructive example inspired by the latter. Suppose that the initial state is of the form
\begin{align}
\ket{\psi_0}=\left\{\begin{array}{ll}
\frac12\left(\ket{j,-1}_y+\sqrt2\ket{j,0}_y+\ket{j,1}_y\right)&j\text{ integer},\\
&\\
\frac{1}{\sqrt2}\left(\ket{j,-\frac12}_y+\ket{j,\frac12}_y\right)&j\text{ semi-odd}.\\
\end{array}\right.
\end{align}
The following properties are easily checked
\begin{align}
\avr{\hat J_y}&=0,\\
\avr{\hat{J}_y^2}&=\left\{\begin{array}{ll}\frac{1}{\sqrt2}& j\text{ integer},\\
\frac12& j\text{ semi-odd}\end{array}\right.\\
\avr{\hat J_z}&=\left\{\begin{array}{ll}\frac{1}{\sqrt2}\sqrt{j(j+1)}& j\text{ integer},\\
\frac12\sqrt{j(j+1)+\frac14}& j\text{ semi-odd}\end{array}\right.
\end{align}
Hence, for any $j>1/2$ we have that $\ket{\psi_0}$ is a spin-squeezed state characterized by a phase sensitivity
\begin{align}
\Delta\phi_{x}^{\sf spin}=\left\{\begin{array}{ll}\frac{1}{\sqrt{j(j+1)}}&j\text{ integer},\\
\frac{1}{\sqrt{j(j+1)+\frac{1}{4}}}&j\text{ semi-odd}.\label{deltaphixesempio}
\end{array}\right.
\end{align}
Clearly, the estimation of $\phi_y$ can be treated along the same lines and yields analogous results.
Recalling the definition of the {\it effective} phase sensitivities relevant for the sequential estimation case [Eqs.~\eqref{eq:effps} and \eqref{eq:phasesens}], we can write down the total phase sensitivity for our sequential estimation strategy based on spin measurements:
\begin{align}
\Delta\Phi_{\sf SE}^{\sf spin}=\left\{\begin{array}{ll}\frac{2}{\sqrt{j(j+1)}}&j\text{ integer},\\
\frac{2}{\sqrt{j(j+1)+\frac{1}{4}}}&j\text{ semi-odd}.
\end{array}\right.\label{eq:SESspin}
\end{align}
This figure has to be compared with the optimal phase sensitivity achievable with a sequential strategy, given in Eq.~\eqref{eq:SES}. One can see that asymptotically the relationship $\Delta\Phi_{\sf SE}^{\sf spin}\sim2\Delta\Phi_{\sf SE}$ holds. Moreover, note that for $j>1/2$ the estimation precision expressed by Eq.~\eqref{eq:SESspin} is always below the standard quantum limit
\begin{align}
	\Delta\Phi_{\sf SE}^{\sf SQL}=\sqrt{\frac2j},
\end{align}
obtained by combining Eqs.~\eqref{eq:effps}, \eqref{eq:phasesens} and $\Delta\phi_x^{\sf SQL}=\Delta\phi_y^{\sf SQL}=1/\sqrt{2j}$. On the other hand, the above discussion shows that the Heisenberg limit $\Delta\phi_x^{\sf HL}=\Delta\phi_y^{\sf HL}=1/{2j}$ is not achievable with spin measurements [this indeed would correspond to achieving Eq.~\eqref{eq:SES}].
\subsection{Two-mode spin-squeezing}
Before concluding, we briefly discuss about the possibility of using a single two-mode spin-squeezed probe state to jointly estimate both phases $\phi_x$ and $\phi_y$. In this case, following the two-mode spin-squeezed property described in Eq. (\ref{eq:twoSS}), the idea would be to prepare a state with reduced fluctuations in the operators $\hat{J}_{x-}$ and $\hat{J}_{y+}$. Indeed, it can be shown that the phase sensitivities obtained by estimating $\phi_x,\phi_y$ through the measurement of respectively $\hat{J}_{y+}$ and $\hat{J}_{x-}$ are given by $\Delta\phi_x\approx\Delta{\hat J_{y+}}/|\avr{\hat J_{z+}}|$ and $\Delta\phi_y\approx\Delta{\hat J_{x-}}/|\avr{\hat J_{z+}}|$. This situation presents some analogies with the scheme presented in Ref.~\cite{DispEst}, where a two-parameter displacement estimation is preformed via a bosonic two-mode squeezed state. However, in that case one is able to associate the two parameters to two commuting observables (at the cost of adding some extra noise), which can then be measured simultaneously. Here, the two observables $\hat J_{x-}^{\sf out}$ and $\hat J_{y+}^{\sf out}$ do not commute in general, although they might commute {\it on average} depending on the probe state of choice. In such a case the question of simultaneous measurement of the two spin operators becomes nontrivial \cite{ozawa}, and goes beyond the scopes of this paper. One may then consider a measurement strategy which is motivated by simplicity: given $M$ copies of the probe state, on $M/2$ of those we perform the measurement of $\hat{J}_{x_{1,2}}$, while the operators $\hat{J}_{y_{1,2}}$ are measured on the remaining $M/2$ copies. Then, $M/2$ values for $\hat{J}_{x-}$ and $\hat{J}_{y+}$ are obtained respectively by subtracting or summing the experimental outcomes. However, it can be seen that this is a `LOCC' measurement strategy \cite{Ballester04}, i.e. it can be simulated by preparing an appropriate ensemble of single-mode spin-$j$ states; this is then equivalent to adopting a `sequential strategy' as in Section.~\ref{s:seq}, with the additional restriction of using the same probe state for both ensembles. These observations suggest that the use of two-mode spin-squeezed probe states combined with spin measurements, which may look appealing by drawing an analogy between our problem and the displacement estimation in bosonic systems \cite{DispEst}, yields in fact no advantage as compared to a simple sequential strategy with single-mode spin-squeezed states.
\begin{figure}[h!]
\includegraphics[width=0.95\columnwidth]{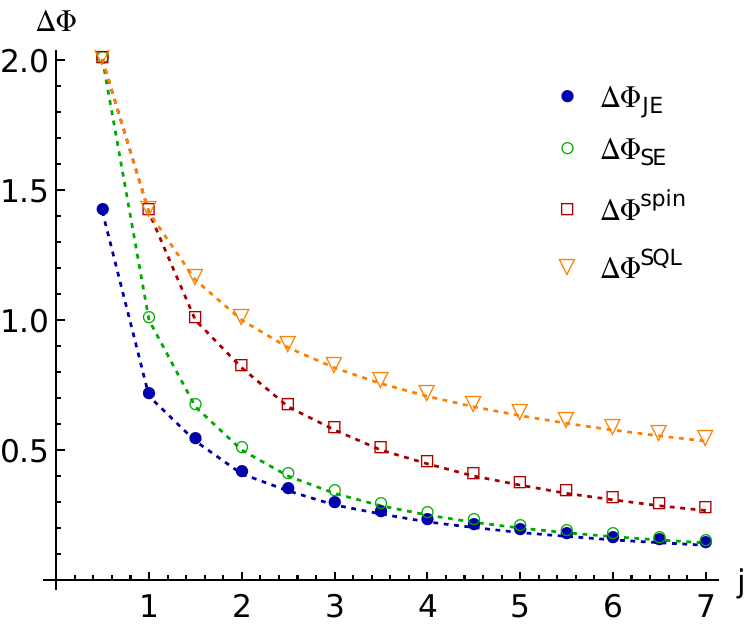}
\caption{Phase sensitivities as a function of the spin-dimension $j$ for the different strategies considered: $\Delta\Phi_{\sf JE}$: joint estimation strategy; $\Delta\Phi_{\sf SE}$: sequential estimation strategy; $\Delta\Phi_{\sf SE}^{\sf spin}$: sequential estimation strategy with spin measurements; $\Delta\Phi_{\sf SE}^{\sf SQL}$: standard quantum limit. 
\label{f:plotscalings}}
\end{figure}

\section{Conclusions}\label{s:concl}
In this paper we have studied the quantum estimation of a two-phase infinitesimal spin rotation. By adopting the standard quantum Cram\'er-Rao bound as a figure of merit, we have derived in closed form the ultimate limits to the estimation precision for both joint and sequential estimation of the two parameters. The results are summarised in Fig. \ref{f:plotscalings}, where the different phase sensitivities are plotted as a function of the spin-dimension $j$. We observed that the joint estimation gives in general a better estimation precision and that in both cases, the asymptotic scaling of the total phase sensitivity with the spin number is $\Delta\Phi\approx1/j$, corresponding to the so-called Heisenberg limit. Then, we have restricted our attention to the precision achievable by adopting an estimation strategy based on spin measurements. We have shown that spin-squeezed states can be employed to beat the standard quantum limit $\Delta\Phi=\sqrt{2/j}$, and we have presented a constructive example of a spin squeezed probe state 
achieving the scaling $\Delta\Phi\approx2/j$.

\section*{Acknowledgments}
CV is thankful to the CQD theory group at Imperial College London for the warm hospitality. TT acknowledges support from the NPRP 4-554-1-084 from Qatar National Research Fund. MGG acknowledges a fellowship support from UK EPSRC (grant EP/I026436/1). 
The authors are very grateful to M. S. Kim for many fruitful discussions.


\begin{thebibliography}{99}
\bibitem{Giovanetti11} V. Giovannetti, S. Lloyd and L. Maccone, Nat. Photonics {\bf 5}, 222 (2011).
\bibitem{Hel} C. W. Helstrom and R. S. Kennedy, IEEE Trans. Inf. Theory {\bf 20}, 16 (1974).
\bibitem{BrauCaves} S. Braunstein and C. Caves, Phys. Rev. Lett. {\bf 72}, 3439 (1994);
S. Braunstein, C. Caves and G. Milburn, Ann. Phys. {\bf 247}, 135 (1996).
\bibitem{paris} M. G. A. Paris, Int. J. Quant. Inf. {\bf 7}, 125 (2009).
\bibitem{Yuen} H. P. Yuen and M. Lax, IEEE Trans. Inf. Theory {\bf IT-19}, 740 (1973).
\bibitem{Belav} V. P. Belavkin, Theoret. Math. Phys. {\bf 3}, 316 (1976).
\bibitem{Holevo} A. S. Holevo, {\em Probabilistic and statistical aspects of quantum
theory} (North-Holland, Amsterdam, 1982).
\bibitem{Naga89} H. Nagaoka, IEICE Technical Report {\bf IT89-42}, 9-14 (1989).
\bibitem{Fuji94} A. Fujiwara, METR 94-9 (1994).
\bibitem{Ballester04} M. A. Ballester, Phys. Rev. A {\bf 69}, 022303 (2004).
\bibitem{cazzumoto} K. Matsumoto, J. Phys. A {\bf35}, 3111 (2002).
%
\bibitem{Chi06} G. Chiribella, G. M. D'Ariano and M. F. Sacchi, J. Phys. A: Math Gen {\bf 39}, 2127 (2006).
\bibitem{You09} K. C. Young, M. Sarovar, R. Kosut and K. B. Whaley, Phys. Rev. A {\bf 79}, 062301 (2009).
\bibitem{Wat10} Y. Watanabe, T. Sagawa and M. Ueda, Phys. Rev. Lett. {\bf 104}, 020401 (2010).
\bibitem{MonIllu01} A. Monras and F. Illuminati, Phys. Rev. A {\bf 81}, 062326 (2010).
\bibitem{MonIllu02} A. Monras and F. Illuminati, Phys. Rev. A {\bf 83}, 012315 (2011).
\bibitem{Cro12} P. J. D. Crowley, A. Datta, M. Barbieri and I. A. Walmsley, arXiv:1206.0043 [quant-ph].
\bibitem{DispEst} M. G. Genoni, M. G. A. Paris, G. Adesso, H. Nha, P. L. Knight and M. S. Kim, 
Phys. Rev. A {\bf 87}, 012107 (2013).
\bibitem{Kahn07} J. Kahn, Phys. Rev. A {\bf 75}, 022326 (2007).
\bibitem{Bisio10} A. Bisio, G. Chiribella, G. M. D'Ariano and P. Perinotti, Phys. Rev. A {\bf 82}, 062305 (2010).
%
\bibitem{kita} M. Kitagawa, M. Ueda, Phys. Rev. A {\bf 47}, 5138 (1993).
\bibitem{vino} D. J. Wineland, J. J.Bollinger, W. M. Itano, F. L. Moore, D. J. Heinzen, Phys. Rev. A {\bf46} R6797 (1992).
\bibitem{entanglo2} N. Bigelow, Nature {\bf 409}, 27 (2001).
\bibitem{precision} A. D. Cronin, J. Schmiedmayer, D. E. Pritchard, Rev. Mod. Phys. {\bf 81}, 1051 (2009).
\bibitem{ramsey} D. J. Wineland, J. J. Bollinger, W. M. Itano, D. J. Heinzen, Phys. Rev. A {\bf 50}, 67 (1994); 
J. Bollinger, W. Itano, D. Wineland, D. Heinzen, Phys. Rev. A {\bf 54} R4649 (1996).
\bibitem{mignot} D. Budker and M. Romalis, Nature Phys. {\bf3}, 227 (2007); W. Wasilewski, K. Jensen, H. Krauter, J. J. Renema, M. V. Balabas, and E. S. Polzik, Phys. Rev. Lett. {\bf 104}, 133601 (2010).
\bibitem{grave} J. N. Hollenhorst, Phys. Rev, D {\bf 19}, 1669 (1979).
\bibitem{busone} D. F. Walls and G. J. Milburn, \textit{Quantum Optics}, Springer, Berlin (2008).
\bibitem{phaseMon} A. Monras, Phys. Rev. A {\bf 73}, 033821 (2006).
\bibitem{Gaiba} R. Gaiba and M. G. A. Paris, Phys. Lett. A {\bf 373}, 934 (2009).
\bibitem{lossMon} A. Monras and M. G. A. Paris, Phys. Rev. Lett. {\bf 98}, 160401 (2007).
\bibitem{carmenKerr} M. G. Genoni, C. Invernizzi and M. G. A. Paris, Phys. Rev. A {\bf 80}, 033842 (2009).
\bibitem{OliBayes} S. Olivares and M. G. A. Paris, J. Phys. B {\bf 42}, 055506 (2009).
\bibitem{PhestCV} M. G. Genoni, S. Olivares and M. G. A. Paris, Phys. Rev. Lett. {\bf 106}, 153603 (2011).
\bibitem{DispMatteo} M. D'Ariano, P. Lo Presti and M. G. A. Paris, Phys. Rev. Lett. {\bf 87}, 270404 (2001).
\bibitem{Kuz00} A. Kuzmich and E. S. Polzik, Phys. Rev. Lett. {\bf 85}, 5639 (2000).
\bibitem{Jul01} B. Julsgaard, A. Kozhekin and E. S. Polzik, Nature {\bf 413}, 400 (2001).
\bibitem{Ber02PRA} D. W. Berry and B. C. Sanders, Phys. Rev. A {\bf 66}, 012313 (2002).
\bibitem{Ber02NJP} D. W. Berry and B. C. Sanders, New. J. Phys. {\bf 4}, 8 (2002).
\bibitem{Ray03} M. G. Raymer, A. C. Funk, B. C. Sanders and H. De Guise, Phys. Rev. A {\bf 67}, 052104 (2003).
\bibitem{Ber05} D. W. Berry and B. C. Sanders, J. Phys. A: Math Gen {\bf 38}, L205 (2005).
\bibitem{yurke} B. Yurke, Phys. Rev. Lett. {\bf56}, 1515 (1986).
\bibitem{ozawa} M. Ozawa, Phys. Lett. A {\bf320}, 367 (2004).
\end{thebibliography}
\end{document}